\documentclass[conference]{IEEEtran}
\IEEEoverridecommandlockouts

\usepackage[inline]{enumitem}
\usepackage{cite}
\usepackage{amsmath,amssymb,amsfonts}
\usepackage{algorithmic}
\usepackage{graphicx}
\usepackage{textcomp}
\usepackage{xcolor}
\usepackage{url}
\usepackage{tabularx}
\usepackage{longtable}
\usepackage{xurl}
\usepackage[hidelinks]{hyperref}
\usepackage[table]{xcolor}
\usepackage{graphicx}
\usepackage{booktabs}
\usepackage{colortbl}
\usepackage{multirow}
\usepackage{array}
\usepackage{booktabs,makecell}
\usepackage{pifont}
\usepackage{csquotes} 
\usepackage{bm} 
\usepackage{comment}
\definecolor{codegray}{rgb}{0.92, 0.92, 0.92}
\usepackage{tablefootnote}

\usepackage{threeparttable}

\definecolor{darkgreen}{RGB}{0,100,0}

\definecolor{darkorange}{RGB}{204,85,0}

\usepackage{listings}
\usepackage{xcolor}

\newcommand{\authornote}[1]{%
  \begingroup
  \renewcommand{\thefootnote}{}%
  \footnotetext{#1}%
  \addtocounter{footnote}{0}%
  \endgroup
}

\lstdefinelanguage{ITPProof}{
    keywords={ have, exact,  apply ,try, done,admit, by , rewrite,  move},
    keywordstyle=\color[rgb]{0.0, 0.35, 0.65}\bfseries, 
    morekeywords=[2]{Context,Variable,Definition,Hypothesis, Lemma,Proof,Qed,Admitted},
   keywordstyle=[2]\color[rgb]{0.9, 0.25, 0.05}\bfseries,
    morekeywords=[3]{forall},
    keywordstyle=[3]\color[rgb]{0.0, 0.45, 0.1}\bfseries,
    sensitive=true,
    morecomment=[l]{--}, 
    commentstyle=\color{gray}\itshape,
    literate={->}{{{\ensuremath{\rightarrow}}}}2
}

\newcommand{\ours}{{ProsaBuddy}}

\def\BibTeX{{\rm B\kern-.05em{\sc i\kern-.025em b}\kern-.08em
    T\kern-.1667em\lower.7ex\hbox{E}\kern-.125emX}}
\begin{document}

\title{ProsaBuddy: Assisting Mechanized Real-Time Schedulability Analysis with LLM-based Agents}

\newcommand{\correspondingauthor}[1]{%
  \begingroup
  \renewcommand{\thefootnote}{}
  \footnotetext{#1}
  \endgroup
}

\author{
\IEEEauthorblockN{
Junyi Liu\textsuperscript{1,2},
Tianchi Ren\textsuperscript{1},
Fei Guan\textsuperscript{3},
Xu Jiang\textsuperscript{4},
Zhe Jiang\textsuperscript{5},
Wang Yi\textsuperscript{6},
Nan Guan\textsuperscript{1}
}

\IEEEauthorblockA{
~~\\
\textsuperscript{1}City University of Hong Kong,
\textsuperscript{2}Northeastern University (China),
\textsuperscript{3}Northeast Forestry University,\\
\textsuperscript{4}University of Electronic Science and Technology of China,
\textsuperscript{5}Southeast University,
\textsuperscript{6}Uppsala University
}
}

\maketitle

\authornote{Junyi Liu and Tianchi Ren contributed equally to this work.\\
Corresponding author: Nan Guan 
\texttt{\string<nanguan@cityu.edu.hk\string>}
}
\begin{abstract}
Rigorous schedulability analysis is essential for the design of hard real-time systems, yet errors in pen-and-paper proofs threaten the safety of critical applications. The Prosa initiative addresses this by offering a foundation for building machine-checkable schedulability analysis proofs in the Rocq proof assistant. However, the substantial time and expertise required to construct such proofs remain a major barrier for wider adoption of Prosa. This work presents ProsaBuddy, an LLM-based agent system designed to lower the effort needed to develop mechanized real-time schedulability proofs. ProsaBuddy employs a ReAct loop with retrieval over the Prosa codebase, access to Rocq tools and optional human-written hints. It uses a subgoal-delegation architecture, decomposing a lemma into subgoals and dispatches them to subagents for proof. 
We evaluate ProsaBuddy on a mini benchmark drawn from real-time scheduling literature. Experiment results show that ProsaBuddy significantly outperforms state-of-the-art LLM-based Rocq automated proving agent systems and a general coding agent OpenCode. 
\end{abstract}


\section{Introduction}
In hard real-time systems, rigorous real-time performance analysis is essential to guarantee the system's timing behavior to consistently meet specified constraints. Researchers in this field typically develop schedulability analysis theorems that provide efficient methods for evaluating real-time performance. The development of such analyses has traditionally relied on human effort and has been conducted in a pen-and-paper manner. However, numerous recent reports have revealed errors in schedulability analyses documented in literatures \cite{erratasuspend,counterexampleEDF,manysuspensions,refuteprob}. Since real-time systems underpin safety-critical applications, such analysis errors can lead to disastrous consequences.

Prosa \cite{prosa} is an initiative that aims to fundamentally address this problem. Prosa provides a repository of definitions and proofs for mechanized (i.e., machine-checkable) real-time schedulability analysis. It is built and verified using Rocq (formerly known as Coq)~\cite{coq}, a mature interactive theorem prover. Prosa builds dedicated infrastructure and methodological frameworks to enable detailed, reliable, yet readable real-time schedulability analysis. Widespread adoption of the Prosa approach within the real-time systems research community would significantly elevate the rigor, credibility, and foundational robustness of theoretical developments in the field.

Despite its potential, a practical barrier prevents Prosa from achieving broader adoption: developing proofs in Rocq is very time-consuming. Although Prosa has  features that enhance its accessibility to real-time systems researchers, such as prioritizing readability and adhering to established proof conventions within the community, the overhead of proof development remains substantial. According to our own experience, proving lemmas in a real-time scheduling paper in Rocq based on Prosa may require several weeks, even for those with moderate familiarity with Prosa and Rocq. For newcomers without prior experience, it is even more demanding. This substantial effort is likely a primary factor discouraging researchers from adopting the Prosa approach in their own work.

In this work, we present \ours, an LLM-based agent system designed to lower the effort barrier to adopting the Prosa approach for developing mechanized schedulability proofs. While LLMs have demonstrated impressive automated theorem proving  capabilities in domains such as pure mathematics, their application to theorem proving in real-time scheduling remains entirely unexplored. Although real-time scheduling lemmas can broadly be regarded as mathematical problems, they exhibit unique features that make them significantly more challenging to formalize and prove automatically than the pure-math problems in standard automated theorem proving benchmarks, as will be discussed in Section \ref{ss:overview}.

We build \ours\ as an agent system that relies on inference‑time scaling through an intensive ReAct (Reasoning \& Acting) loop~\cite{react}. It combines retrieval over the Prosa codebase to supply necessary definitions and proof patterns, access to various Rocq tools which provide rich feedback information, and optional human‑written proof hints that provide high‑level guidance. \ours\ adopts a subgoal‑delegation architecture: a primary agent decomposes the lemma into smaller subgoals and dispatches them to subagents, each receiving a structured task bundle that includes the subgoal, context, and a proof state snapshot. We enhance the agents with decomposition guidelines and domain‑specific proving skills, improving their ability to navigate the deep proof chains.

{We conducted experiments to evaluate the performance of \ours, comparing it with state-of-the-art LLM-based Rocq automated theorem proving systems, as well as a general-purpose coding agent OpenCode~\cite{opencode}.}
The comparison is conducted on a mini-benchmark of $24$ lemmas that we built from the real-time scheduling literature; none of them appear in the existing Prosa library.
Experimental results show that \ours \ significantly outperforms all competing systems.
We further extend the evaluation to a larger set of lemmas taken directly from the Prosa library, and the results confirm the superiority of \ours.
Ablation studies reveal that these gains arise from the combined effect of several design choices in \ours.
Additionally, we assess the impact of different backbone LLMs on the performance of \ours.

Overall, \ours\ successfully proves non-trivial schedulability analysis lemmas, including some whose Rocq proofs require hundreds of tactics. Nevertheless, we do not currently position \ours\ as a replacement for human researchers or as capable of proving such lemmas entirely independently. Instead, \ours\ is designed to liberate researchers from the tedious low-level details of lengthy Rocq proofs, saving them time to focus on developing high-level proof ideas. However, given the rapid development of LLMs, a vast open space remains for exploring whether \ours\ could, in the future, assist real-time scheduling research in more advanced ways, for instance, by independently proving open problems or even automatically developing new theories.

\section{Background and Related Work}\label{background}
\subsection{Rocq}

Rocq (formerly known as Coq) is a proof assistant based on dependent type theory, allowing users to write mathematical definitions, executable algorithms, and prove properties about them interactively. Following the project's renaming, we will refer to Rocq exclusively when introducing \ours.
One key difference between human‑written and Rocq proofs is the level of detail: a simple intuition often requires many explicit steps. For example, in the example of Figure \ref{fig:scheduled_not_completed}, the lemma states that a job scheduled at time $t$ cannot be completed by $t$. For humans, this is trivial. In Rocq, however, the proof needs many steps. So developing Rocq proofs is effort‑demanding.

Rocq does not draw a fundamental distinction between \texttt{Lemma}, \texttt{Theorem}, \texttt{Fact}, or similar keywords – they all introduce a goal to be proved. In this paper, we will use the term \emph{lemma} uniformly for any such goal to prove, regardless of which keyword was used in its declaration.
A lemma statement in Rocq typically introduces variables, definitions, hypotheses, and a lemma statement. 
The proof is enclosed by \texttt{Proof.} and \texttt{Qed.}, and the user guides the proof by applying \emph{tactics}, which are commands that transform the current proof state (e.g., those marked in blue in 
Figure~\ref{fig:scheduled_not_completed}). Rocq provides a rich set of tactics, ranging from basic, primitive tactics that manipulate the proof state directly (e.g., \texttt{move}, \texttt{rewrite}, \texttt{apply}, \texttt{exact}) to powerful automation tactics that can search for proofs or solve decidable fragments automatically (e.g., \texttt{auto}, \texttt{eauto}, \texttt{lia}, \texttt{ring}).

In the presentation of this paper, we deliberately minimize the reader’s dependence on understanding the syntax and semantics of Rocq. Our goal is to let readers who are not familiar with Rocq still grasp the key ideas, even if they skip some details. For some places that are key to understanding our methodology, we will introduce some tactics locally. For a comprehensive introduction to tactics mentioned in this paper, we refer the reader to the Rocq reference manual \cite{coq_manual}.

\begin{figure}
\tiny
\begin{lstlisting}[
 basicstyle=\fontsize{6}{7}\selectfont\ttfamily,
breaklines=false,
  columns=fullflexible,
  keepspaces=true,
  escapeinside={(*@}{@*)}
]
Context {Job: eqType}.
Variable job_arrival: Job -> time.
Variable job_cost: Job -> time.
Variable sched: schedule Job.
Context {Task: eqType}. 
Variable job_task: Job -> Task. 
Definition scheduled_at (j: Job) (t: time) := sched t == Some j.
Definition service (j: Job) (t: time) :=
\sum_(0 <= t' < t) scheduled_at j t'.
Definition completed_by (j: Job) (t: time) := 
        job_cost j <= service j t.
Hypothesis H_service_leq_job_cost:
  forall j t, service j t <= job_cost j.
Lemma scheduled_implies_not_completed:
  forall j t, scheduled_at j t -> ~~ completed_by j t.
Proof.
  move => j t.
  rewrite /completed_by. move=> H_sched. rewrite -ltnNge.
  have H_comp := H_service_leq_job_cost j (t + 1).
  rewrite /service in H_comp. rewrite addn1 in H_comp.
  rewrite big_nat_recr //= in H_comp.
  rewrite H_sched /= in H_comp. rewrite /service.
  by rewrite addn1 in H_comp.
Qed.
\end{lstlisting}
\caption{Example of a simple schedulability analysis lemma in Rocq}
\label{fig:scheduled_not_completed}
\end{figure}

\subsection{Prosa}

Prosa \cite{prosa} is a Rocq-based foundation for verified real-time scheduling theory. It emphasizes readable specifications, explicit assumptions, and proof reuse. Following the SSReflect methodology \cite{ssreflect}, Prosa encodes its specifications as computable boolean predicates, which effectively enables proof automation via reflection.
This clarity also makes Prosa well‑suited for LLM‑based proof automation, as its good readability is informative not only to human users but also to LLM.
Prosa has been extended to verified scheduling infrastructure in several directions in follow‑up work, e.g., sustainability theory and weak sustainability~\cite{strongweaksustainability}, Abstract Response‑Time Analysis (ARTA) based on the busy‑window principle \cite{aRTA}, FIFO response‑time analysis with arbitrary deadlines and arrival curves~\cite{intuition2coq}, and connections between response‑time analysis and network calculus~\cite{linkRandNc}. Furthermore, Prosa serves as the certificate generator for POET~\cite{pOET}, the first foundational response‑time analysis tool that produces fully machine‑checkable Coq/Rocq certificates. Another recent work integrates Prosa with RefinedC~\cite{refinedProsa}, connecting Prosa’s verified response‑time analysis with C implementation verification, thereby enabling end‑to‑end certification of interrupt‑free real‑time schedulers implemented in low‑level code.
Together, these works make Prosa a useful infrastructure for verified real‑time systems. Nevertheless, Prosa currently focuses on a core subset of real‑time scheduling theory.

\subsection{Automated Theorem Proving}

Mechanized proofs are inherently hard to construct. Before the emergence of LLMs, many efforts sought to improve automation for different theorem provers, for instance, via powerful tactics that search for proofs or solve decidable fragments automatically. For Coq/Rocq, SSReflect~\cite{ssreflect}, Mathematical Components~\cite{mathcomp} and CoqEAL~\cite{coqeal} enable reuse of recurring proof patterns through reflection and proof‑oriented computation; SMTCoq~\cite{smtcoq}, smt4coq~\cite{smt4coq} and CoqHammer~\cite{coqhammer} connect Rocq goals to external solvers and reconstruct their proofs inside the kernel; Ltac/Ltac2~\cite{ltac}, Mtac/Mtac2~\cite{mtac,mtac2}, Coq‑Elpi~\cite{coqelpi} and MetaCoq~\cite{metacoq} support domain‑specific tactics and reusable proof infrastructure. ML4PG~\cite{ml4pg}, CoqGym~\cite{coqgym}, Tactician~\cite{tactician}, Proverbot9001~\cite{proverbot9001}, TacTok~\cite{tactok} and ASTactic~\cite{coqgym} explored proof‑pattern mining, tactic prediction, local proof search, and proof‑step prediction.
These techniques substantially reduce routine proof effort, but they do not fully eliminate the high demand for expert intervention. Large mechanized developments still require experts to decompose complex goals, formulate auxiliary lemmas, choose suitable abstractions, and coordinate induction arguments.

\begin{table}[t]
\centering
\footnotesize
\setlength{\tabcolsep}{2.5pt}
\renewcommand{\arraystretch}{0.95}
\caption{LLM-based automated Coq/Rocq theorem proving systems}
\label{tab:llm_proof_automation}
\begin{tabular}{cccccc}
\toprule
Name& Source & Year & Prover & Availability  \\
\midrule
CoqPilot & \cite{coqpilot}
& 2024 & Coq & Yes 
 \\

PALM & \cite{palm}
& 2024 & Coq & Yes  \\

COPRA & \cite{copra}
& 2024 & Coq & Yes  \\

Liu etc. & \cite{liu2024llm}
& 2024 & Coq & No  \\

Lu etc. & \cite{lu2025clarifying}
& 2025 & Coq & No  \\

Rango & \cite{rango}
& 2025 & Coq & Yes  \\

ADAPT & \cite{adaptive}
& 2025 & Coq & No  \\

RocqStar & \cite{rocqstar}
& 2026 & Rocq & Yes \\

Cobblestone  & \cite{cobblestone}
& 2026 & Coq & Yes  \\

AutoRocq & \cite{autorocq}
& 2026 & Rocq & Yes  \\

ReCent-Prover & \cite{recentprover}
& 2026 & Rocq & No  \\

QUARRY & \cite{quarry}
& 2026 & Rocq & No  \\
\bottomrule
\end{tabular}
\label{t:list}
\end{table}

LLMs have recently been used for automated theorem proving across several proof assistants. 
Existing work has been done for different theorem provers, including Coq/Rocq~\cite{coqpilot,copra,palm,rango,rocqstar,recentprover}, Lean~\cite{autoformalization_llm,draftsketch,rautoformalizer,leandojo,llemma,legoprover,deepseekprover,lin2025leanstar,leancopilot,goedelprover,hilbert} and Isabelle/HOL~\cite{thor,baldur,poetry}. 
In the following, we focus on the work for Coq/Rocq, and refer to recent survey papers \cite{autoformalization_survey,llm_atp_survey} for a more comprehensive overview of the field.
CoqPilot~\cite{coqpilot} provides a Coq-oriented assistant implemented as a VS Code extension, where users mark proof holes with \texttt{admit}, configure completion services, and invoke the tool inside an open Coq file to generate, check, and substitute candidate proofs through \texttt{coq-lsp} during interactive editing. COPRA~\cite{copra} formulates proof search as an in-context interactive process where the model proposes tactics and observes proof-state feedback. PALM~\cite{palm} uses a generate-then-repair architecture to analyze and repair low-level proof errors. More recent systems combine proof generation with retrieval and multi-step interaction. Rango~\cite{rango} retrieves relevant premises and similar proofs from the current Coq project. RocqStar~\cite{rocqstar}  and ReCent-Prover~\cite{recentprover} further introduce planning, reflection, and strategy-aware retrieval into Rocq proof generation. 
{Recent systems also enrich proof context and refine failed
proof attempts. Lu et al.\cite{lu2025clarifying} use selective
concept unfolding and separate proof planning from tactic generation.
Liu et al.\cite{liu2024llm} explain terms in proof obligations and
repair incomplete tactic parameters. ADAPT \cite{adaptive} selects
among different refinement strategies after a failed proof attempt.
QUARRY \cite{quarry} ranks candidate sublemma decompositions by their
estimated CoqHammer solvability. Cobblestone\cite{cobblestone}
retains Coq-accepted subproofs and recursively processes failed parts,
while AutoRocq\cite{autorocq} combines proof-tree context, Rocq
feedback, and previous proof attempts for software verification.} {Table~\ref{t:list} lists existing
LLM-based automated Coq/Rocq theorem proving systems.}

Overall, this body of work shows a shift from isolated tactic prediction to systems combining proof synthesis, retrieval, prover feedback, repair and agentic control. However, proofs in real-time scheduling often rely on recurring semantic structures, including scheduling assumptions, workload and processor models, invariants, response-time bounds, rewriting conventions, and sequences of lemmas. Existing methods treat reusable knowledge as retrieved premises, similar proofs, or short-term proof-state contexts. They do not make domain-specific proof procedures explicit as reusable skills with applicability conditions, required assumptions, expected effects and adaptation strategies. This leaves a gap between general LLM-based proof generation and domain-aware proof automation.

\section{ProsaBuddy}\label{s:prosabuddy}
\subsection{Overview}
\label{ss:overview}

Before presenting our approach, we first discuss the major differences between the problem tackled in this paper, automated theorem proving for real‑time schedulability analysis, and the more commonly studied application of LLM‑based automated theorem proving to pure mathematics.

The theorem-proving capability of LLMs is obtained via pre‑training on large‑scale corpora such as Mathlib~\cite{mathlib}. In contrast, public data for mechanized schedulability analysis is very limited. The Prosa library is the main available resource, but its volume is orders of magnitude smaller than the corpora for mathematics. This scarcity fundamentally constrains the ability of LLMs to learn proof patterns specific to the schedulability analysis domain.

The schedulability analysis lemmas are substantially more complex than those found in mathematics benchmarks. They are built upon a complex hierarchy of concepts, requiring many intricate definitions that often recursively depend on one another across multiple levels, along with numerous hypotheses.
For example, in the PutnamBench suite~\cite{putnambench} for evaluating LLM-based mathematics theorem proving, the description of a lemma spans no more than $20$ lines. In contrast, the complete description of a schedulability analysis lemma in Prosa, including all definitions, variables, assumptions, and the lemma itself, often extends to several hundred lines. This increased length and conceptual depth impose a significantly heavier burden on LLM's context handling and reasoning capabilities.
Beyond the complexity of individual lemmas, Prosa has a highly structured architecture that promotes lemma reuse across different task models. Whether an existing lemma can be applied to prove a target lemma depends on whether its preconditions hold, and those preconditions, in turn, rely on further preconditions, ultimately tracing back to the specific assumptions of the task model under consideration. 
This unique feature has introduced extra difficulty in the proving.
We will further elaborate on this point in Section \ref{subsec:proving_skills}.

To address the above challenges, \ours\ adopts several complementary strategies. First, because pre-training  cannot supply sufficient domain knowledge, we rely on inference-time scaling: the LLM repeatedly attempts the proof, using feedback from external tools. 
Second, we leverage the existing Prosa library to retrieve relevant definitions and lemmas, and to learn proving techniques by imitating how Prosa handles similar problems. 
This demands a powerful retrieval capability capable of accurately locating useful information within Prosa. Third, we incorporate optional human-provided proof hints, which help the agent navigate deep proof chains that would otherwise be difficult to resolve automatically.

For these considerations, we build \ours\ on top of OpenCode~\cite{opencode} rather than constructing an agent system from scratch. OpenCode provides a solid foundation that satisfies the requirements outlined above. In OpenCode, an agent follows an iterative ReAct (Reasoning 
\& Acting) loop \cite{react}: it queries the LLM, uses the reasoning in the model's output to execute external tools, and then appends the feedback to the conversation context for the next query, repeating this process until the task is resolved. OpenCode offers powerful code search capabilities well-suited for retrieval over the Prosa library and allows users to inject human-written proof hints into the agent's prompt easily. By building on OpenCode, we inherit a mature agent foundation and can focus our efforts on developing high-level capabilities tailored to the needs of schedulability analysis lemma proving.

\begin{figure}
    \centering
    \includegraphics[width=0.9\columnwidth, trim=1 1 1 1, clip]{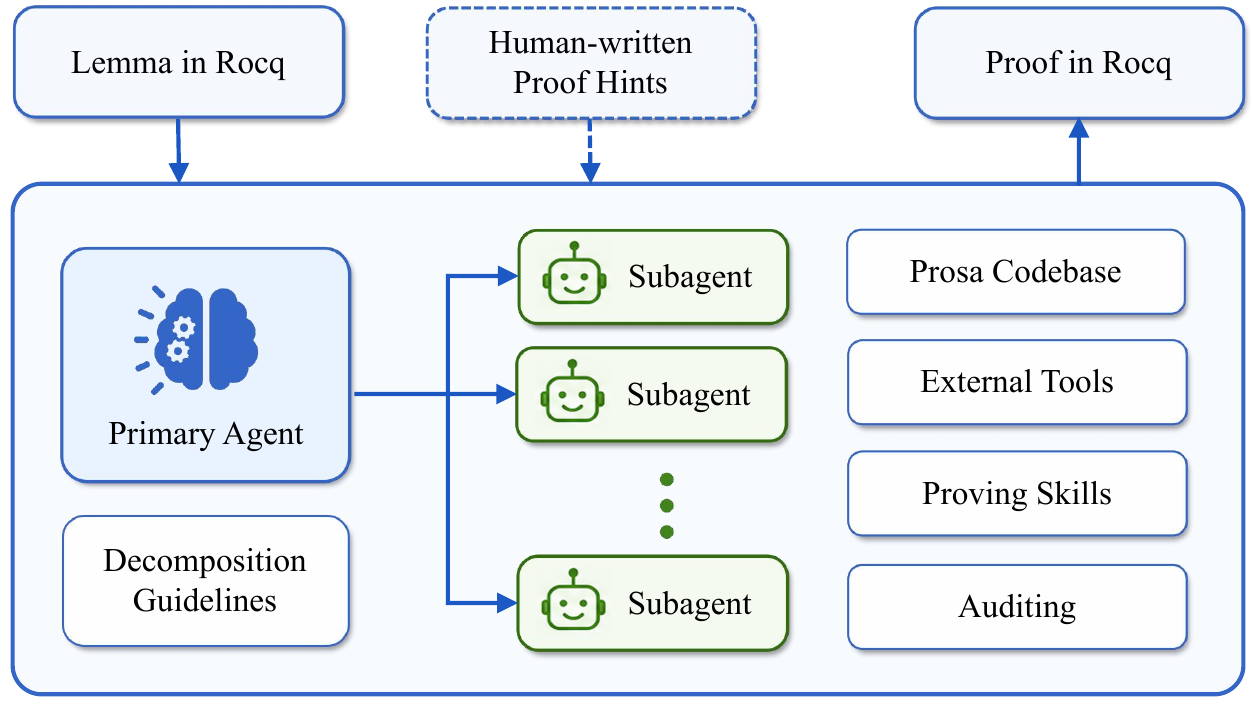}
    \caption{Overview of \ours }
    \label{fig:prosabuddy_arch}
 \vspace{-5pt}
\end{figure}

Figure \ref{fig:prosabuddy_arch} shows the major components of \ours. The system takes as input a lemma formalized in Rocq, along with \emph{optional} human-written proof hints. The output is a verified proof in Rocq, or a failure declaration. For proof construction, \ours\ uses LLM-based agents in a subgoal-delegation architecture. The primary agent decomposes the lemma into smaller, independent subgoals and dispatches them to specialized subagents. The subagent leverages OpenCode's existing infrastructure, which already supports launching subagents. When launching a subagent, the system does not simply forward the original prompt. Instead, it builds a structured task bundle that includes the subgoal to prove, the necessary context, and a snapshot of the current proof state, for the subagent to solve the assigned subgoal independently.
We detail the lemma decomposition process in Section~\ref{subsec:lemma_decomposition}.
The agents retrieve relevant information from the Prosa codebase using OpenCode's standard file-reading and code-search tools, such as file browsing and grep-like search. 
Beyond these standard capabilities, we enable \ours\ to invoke Rocq tools directly to get rich feedback information for its ReAct loop.
We elaborate on these external tools in Section~\ref{subsec:external_tools}.
We also equip the agents with a collection of proving skills that target common challenges in the proving procedure, which will be elaborated in Section~\ref{subsec:proving_skills}.
Last but not least, the proof submitted 
by the agents must go through auditing 
before it is finally accepted as a valid proof of the target lemma, which will be introduced in Section~\ref{subsec:auditing}.

\subsection{Lemma Decomposition}
\label{subsec:lemma_decomposition}
\begin{table*}[htbp]
\centering
\caption{Summary of Decomposition Guidelines}
\label{tab:guiding_principles_for_decomposition}
\scriptsize
\setlength{\tabcolsep}{4pt}
\renewcommand{\arraystretch}{1.2}
\begin{tabular}{
    >{\centering\arraybackslash}p{0.15\columnwidth}
    >{\raggedright\arraybackslash}p{1.83\columnwidth}
}
\toprule
\multicolumn{1}{c}{\textbf{Name}} &
\multicolumn{1}{c}{\textbf{Brief Description}} \\
\midrule
\textbf{Reshaping} 
& If an expression is mathematically useful but its exact form does not match what a downstream lemma requires, we should create a {subgoal} dedicated to reshaping the expression into the exact needed form.
\textbf{Example:} If the context contains $\texttt{size (filter P xs)} \le 1$, a useful subgoal is to reshape it into $\texttt{count P xs} \le 1$, so that a counting lemma can be applied directly.
 \\
\midrule
\textbf{Witnessing} &  An object may already be known to exist,
but it is still hidden inside an existential statement or an informal paper description.
Then the proof should create a subgoal to make that object explicit and record the facts that come with it before using it in later steps.
\textbf{Example:} When the context contains an existential fact about a job arriving in an interval \texttt{I} and belonging to task \texttt{tsk}, a useful subgoal is to extract a concrete witness \texttt{j} together with facts such as \texttt{arrives\_in j I} and \texttt{job\_task j = tsk}.
 \\
\midrule
\textbf{Library Matching} &  Prosa, MathComp, or local lemma may already be the right tool, but the current proof state is not ready for applying it.
Then the proof should create a subgoal
dedicated to preparing that lemma application:
specializing general hypotheses to the current case,
and writing out the required premises
as explicit subgoals to be proved.
\textbf{Example:} Suppose a Prosa lemma can prove the desired conclusion once it is given premises such as
\texttt{job\_task j = tsk}
and
\texttt{arrives\_in j I}.
A useful low-level decomposition is to create separate subgoals for these premises, prove them first, and then apply the library lemma directly.
 \\
\midrule
\textbf{Case Splitting} & A proof may need to be split into cases, but each branch should not be packed into one big proof region.
Then the proof should create a subgoal
to make the case split explicit:
write out the branch condition,
and leave the rest of the branch proof
to later subgoals.
\textbf{Example:} We already know that the next step depends on whether a job \texttt{j}
belongs to the set of jobs arriving in interval \texttt{I},
but we should have a subgoal dedicated to splitting 
\texttt{j}~$\in$~\texttt{jobs\_arrived\_during I},
so that one branch explicitly carries the fact \texttt{j}~$\in$~\texttt{jobs\_arrived\_during I},
and the other branch explicitly carries the fact
\texttt{j}~$\notin$~\texttt{jobs\_arrived\_during I}.
Then the latter proof can handle the two branches separately, instead of hiding the whole case analysis inside one large proof step.
 \\
\midrule
\textbf{Lifting} & When a goal concerns an set (e.g., counting, cardinality, or aggregation) but only per-element implications are available, introduce a subgoal to lift them to the set level. 
\textbf{Example:} 
If each job \texttt{j} that receives service in an interval \texttt{I} must have arrived before the end of~\texttt{I}, a useful subgoal is to lift this fact to a counting inequality:
\texttt{count (fun j => receives\_service\_in I j) jobs} $\le$
\texttt{count (fun j => arrived\_before (end\_of I) j) jobs}.
 \\
\bottomrule
\end{tabular}
\end{table*}

\begin{figure}
    \centering
    \includegraphics[width=1.05\columnwidth, trim=1 1 1 1, clip]{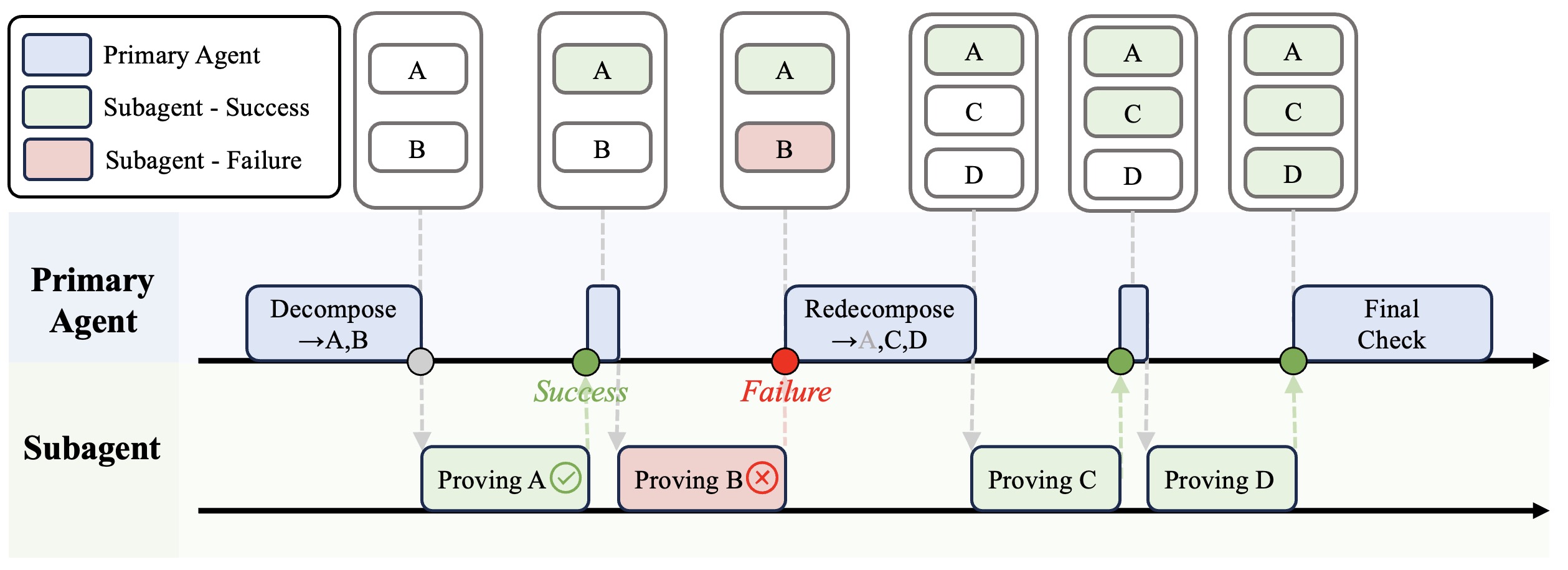}
    \caption{An example of the decomposition-based proving procedure}
    \label{fig:workflow}
 \vspace{-5pt}
\end{figure}

The proving process starts with the primary agent, which first decides whether the lemma to be proved is simple enough to be proven by itself; if so, it attempts to complete the proof directly. If the lemma is complex, it performs lemma decomposition, i.e., breaks the target lemma into several subgoals. Each subgoal is then delegated to a subagent for proof. After launching a subagent, the primary agent suspends and waits for that subagent to return. The subagent returns to the primary agent with either a success or a failure, accompanied by a failure report that explains why it failed and what important piece of information is missing for it to complete the proof of that subgoal. Upon receiving a failure return from a subagent, the primary agent attempts to refine the decomposition. 
When the primary agent refines the decomposition, all subgoals that have already been proved will be kept.
Figure \ref{fig:workflow} shows an illustrative example of the proving process. 
The primary agent first decomposes the target lemma into two subgoals, \texttt{A} and \texttt{B}. The subagent proving \texttt{A} returns with a success, but the one proving \texttt{B} returns with a failure. The primary agent then analyzes the failed subgoal, redo the decomposition, further divides subgoal \texttt{B} into smaller subgoals \texttt{C} and \texttt{D}, and dispatches them to subagents for proof. Once both \texttt{C} and \texttt{D} are successfully proved, the primary agent performs a final recheck to ensure that all generated proofs returned from the subagents compose into a valid proof of the original lemma.

The decomposition by the primary agent is mechanically checked for structural correctness: it guarantees that the original lemma follows from the conjunction of the subgoals. 
Concretely, the agent constructs a \emph{proof skeleton} in which each subgoal is replaced by \texttt{admit}, and 
then verifies that this proof is accepted by the Rocq compiler, 
as illustrated in
Figure~\ref{fig:decomp_example}.
Nested decomposition is permitted (e.g., introducing a \texttt{have} tactic within another \texttt{have}, provided it reflects a genuine sub-step or a structural refinement of the reasoning). 
Intermediate subgoals (those that are further decomposed) are not delegated to subagents; only 
{terminal} subgoals (those that are \emph{not} further decomposed) are delegated to subagents for proof.
In Figure~\ref{fig:decomp_example}, the original lemma \texttt{T} is decomposed into \texttt{H1} and \texttt{H2}, where \texttt{H1} is an intermediate subgoal and \texttt{H2} is terminal. 
{\texttt{H1} is further decomposed into the terminal subgoals \texttt{S1} and \texttt{S2}. The terminal subgoals \texttt{S1}, \texttt{S2} and \texttt{H2} are then dispatched to subagents for proof.}

\begin{figure}
\tiny
\begin{lstlisting}[
 basicstyle=\fontsize{6}{7}\selectfont\ttfamily,
breaklines=false,
  columns=fullflexible,
  keepspaces=true,
  escapeinside={(*@}{@*)}
]
Lemma T (a b c d e : nat) :
  a + c <= b + d.
Proof.
  have H1 : a <= b.
  {
    have S1 : a <= e.
    { admit. }
    have S2 : e <= b.
    { admit. }
    exact: leq_trans S1 S2.
  }
  have H2 : c <= d.
  { admit. }
  exact: leq_add H1 H2.
Admitted.
\end{lstlisting}
\caption{An example of the proof script for lemma decomposition}
\label{fig:decomp_example}
\end{figure}

The primary agent performs lemma decomposition in two stages. First, a high-level decomposition stage reasons about the current context (incorporating any human-written proof hint, if given) to extract fundamental steps that can genuinely advance the proof. Second, a low-level decomposition stage transforms those high-level steps into a proof skeleton directly provable in Rocq. It turns out that the agent is doing relatively well with the high-level decomposition, especially when human-written proof hints are given. However, the agent struggles in the second stage. 
Because the low-level decomposition relies on the backbone LLM's inherent reasoning capability, it is usually not fine-grained enough, and thus the resulting subgoals are difficult for the subagent to prove.
To solve this problem, we further augment the agent with extra guiding principles that are extracted manually from our experience in using Rocq for theorem proving. These principles provide instructions tailored to situations where low-level decomposition is effective (summarized in Table~\ref{tab:guiding_principles_for_decomposition}). Specifically, they target highly localized sub-problems that are largely independent of the surrounding proof, making them ideal for a subagent to solve in a clean, isolated context. The guiding principles are embedded in the main agent’s prompt. The full prompt is considerably more detailed, and Table~\ref{tab:guiding_principles_for_decomposition} gives only a brief overview.



\subsection{Proving Skills}
\label{subsec:proving_skills}

\begin{table*}[htbp]
\centering
\begin{threeparttable}
\caption{Summary of Proving Skills}
\label{tab:proving_skill_summary}
\scriptsize
\setlength{\tabcolsep}{4pt}
\renewcommand{\arraystretch}{1.2}
\begin{tabular}{
    >{\bfseries\raggedright\arraybackslash}p{0.13\textwidth}
    >{\raggedright\arraybackslash}p{0.83\textwidth}
}
\toprule
 \multicolumn{1}{c}{\textbf{Name}} & \multicolumn{1}{c}{\textbf{Brief Description}} \\
\midrule
Small-step \& inspect & Write one tactic at a time and inspect the resulting goal state after each step. This avoids writing long proof scripts without checking intermediate goals, and it helps keep the actual proof state in view. \\
\midrule
 Error priority  & Resolve errors in priority order: syntax or parse errors first, then ill-typed terms, then unsolved goals or tactic failures, since later diagnostics are often consequences of earlier failures. \\
\midrule
Tactic selection & Use simpler and more direct tactics before heavier or more search-based ones, such as \texttt{reflexivity}, \texttt{lia}, \texttt{ring}, \texttt{tauto}, \texttt{congruence}, \texttt{auto}/\texttt{eauto}, and \texttt{decide}/\texttt{native\_decide}. For composite goals, decompose the goal first and automate each component separately. \\
\midrule
 Rewrite discipline & When \texttt{rewrite} fails because the rewritten term appears in dependent types, reorganize  context with commands like \texttt{generalize dependent} or \texttt{subst}. Also, name small helper results with \texttt{assert}, \texttt{pose proof}, or \texttt{set} instead of packing too many steps into one long tactic chain. \\
\midrule
 Cleanup \& final check & After a proof compiles, clean it up, choose deliberately between \texttt{Qed} and \texttt{Defined}, and then actually compile the file with a command such as  \texttt{coqc} to make sure everything still works. A proof is not complete until temporary holes such as \texttt{admit} or \texttt{Admitted.} have been removed. \\
\midrule
 Syntax-driven rewrite & Before each rewrite, check whether the pattern required by the lemma already appears exactly in the current goal. If it does not, first transform the goal into that form instead of forcing the rewrite just because it seems mathematically correct. \\
\midrule
 Local bridge lemmas & Use a small local bridge lemma when the mismatch is specific to the current branch or to a fixed local expression. Extract a generic lemma only when the same normalization gap recurs across proofs. \\
\midrule
 Minimal expansion & When a \texttt{by} fails, remove \texttt{by} and keep the rest exactly the same. Re-run immediately and record both the first revealed error message and the number of focused goals. If expansion produces multiple goals, the primary issue is proof structure. If it reveals a rewrite mismatch, the issue is the current goal shape. If it reveals \texttt{Cannot apply} or \texttt{No assumption}, the issue lies in the lemma preconditions or the local context. \\
\midrule
 Expand ``by'' patterns & Expand compressed forms such as \texttt{by rewrite}, \texttt{by apply:}, \texttt{have \dots by \dots}, and \texttt{case \dots; by \dots} into ordinary proof scripts so that you can see exactly which step fails or which branch is still open. After expanding one \texttt{by} step, fix only that step. Do not guess new tactics before reading the first real error, and shorten it back to \texttt{by} only after the expanded version is confirmed to close the proof correctly. \\
\midrule
 Bool/Prop boundary & If the remaining reasoning is easier to carry out in Prop than in Boolean form, convert the comparison once and finish the branch in Prop, reflecting back only if the final goal is Boolean. \\
\midrule
 Normal form \& bridge order & First decide what form the final result should have: a count expression or a big-operator expression. After choosing that target form, rewrite in one direction: \texttt{indicator sum -> filtered big operator -> count}. Do not switch back and forth between forms. \\
\midrule
 Big-op normalization order & First simplify the innermost sums that are easy to evaluate, such as constant sums or sums of Boolean tests. Then reorganize the outer sum if needed. Only after that should you rewrite repeated addition as multiplication. Try to keep the whole branch in one consistent form instead of repeatedly switching between sums, \texttt{iter}, and multiplication. \\
\midrule
 Count before min & First rewrite the main counting fact into a clear bound on \texttt{count}, such as \texttt{count P s <= k}. Only after that should you handle \texttt{Nat.min}, \texttt{minn}, or the final arithmetic inequality. Do not start the final arithmetic steps while the same quantity is still written in several different counting forms. \\
\midrule
 Failure signals & If \texttt{big\_mkcond} or \texttt{sum1\_count} does not apply, the expression is usually still in the wrong form. A common mistake is to skip the intermediate normalization steps and jump straight from an indicator sum to \texttt{count} or even to a \texttt{min} inequality, which makes both the rewrites and the later arithmetic harder to control. \\
\midrule
 Failure signatures & When rewrites such as \texttt{mul1n}, \texttt{mulnC}, or \texttt{big\_ord\_recr} fail, the expression is usually still in the wrong form. If the branch has become a counting problem involving \texttt{count}, \texttt{sum1\_count}, or \texttt{Nat.min}, switch to the count-normalization skill instead of continuing with generic big-operator rewrites. \\
\midrule
 Goal management & Use this skill when a Coq or Ssreflect proof has drifted into a multi-goal state after a branching tactic or local sublemma, especially when \texttt{coqc} reports ``\texttt{Expected a single focused goal but 2 goals are focused,}'' and the next step should be structural recovery with \texttt{bullets}, \texttt{braces}, and \texttt{explicit branch closure} rather than further mathematical reasoning.\tnote{1} \\
\midrule
 Goal-driven lemma instantiation & This skill helps distinguish between ordinary subgoals that can be proved later and early parameters or prerequisites that Coq must know immediately. It recommends using apply or eapply to let the current goal expose the real remaining obligations, and, if necessary, supplying only the first missing parameter or preconditions instead of writing a long, fully explicit exact term.\\
\bottomrule
\end{tabular}

\scriptsize
\begin{tablenotes}[flushleft]
\item[1]Goal-management errors were the most frequent failures in our experiments.
This is likely because Rocq and newer SSReflect proof modes enforce goal focusing and subproof bookkeeping more explicitly, so compressed proof scripts that worked in older Coq settings may fail in Rocq.
We hypothesize that current LLMs have seen more older Coq scripts than Rocq 9.x scripts, and therefore often generate fragile goal-management patterns that lead to these errors.
\end{tablenotes}

\end{threeparttable}
\end{table*}

We develop proving skills to enhance the subagent's ability to prove delegated subgoals. These skills are generated by using the LLM to analyze the proving traces, using the following process.
First, we select $65$ lemmas from Prosa (only a small fraction of Prosa) as \emph{training lemmas}\footnote{For detailed information on the training lemmas,
see the ProsaBuddy Github repository
\href{https://github.com/ProsaBuddy/RTSS2026}
{\(\langle\)\nolinkurl{https://github.com/ProsaBuddy/RTSS2026}\(\rangle\)}.}.
The selection criteria aim for diversity: we choose lemmas with proofs that vary in tactic count (i.e., difficulty), and we sample from different files and, whenever possible, different subdirectories within the Prosa codebase. 
We use a preliminary version of our system (equipped with external Rocq tools and the Prosa codebase) to attempt to prove these lemmas, and we collect the full traces of the proving procedure. 
Each trace captures the agent's iterative attempts, including failures, error messages, and proof states, until success.
We then fed these traces to an LLM to analyze failure reasons and generate skills. Concretely:
\begin{enumerate}
\item  We use LLM to divide each trace into segments, each handling a local task (e.g., fixing an error in the proof). 

\item For each segment, we use the LLM to identify the problem being solved, analyze its cause, and generate a skill to avoid or resolve it. This produces a pool of proving skill candidates.

\item We use the LLM to merge proving skill candidates that address the same or similar problems.

\item Finally, we manually review the remaining candidates, discarding those that are too general, vague, or lacking useful information, and enhance some of them with human expertise. This results in the final $17$ proving skills as summarized in Table \ref{tab:proving_skill_summary}.
\end{enumerate}

\begin{figure}
\begin{lstlisting}[
 basicstyle=\fontsize{6}{7}\selectfont\ttfamily,
breaklines=false,
  columns=fullflexible,
  keepspaces=true,
  escapeinside={(*@}{@*)}
]
exact (@uniprocessor_response_time_bound_restricted_supply_seq
    Task H (@limited_preemptions_rtc_threshold Task H H1)
    Job H2 H4 H3 (@limited_preemptive_job_model Job H5)
    PState H_uniprocessor_proc_model H_unit_supply_proc_model
    H_consumed_supply_proc_model arr_seq VALID_ARRIVALS
    sched JOBS_FROM_ARRIVAL_SEQUENCE JOBS_MUST_ARRIVE
    COMPLETED_JOBS_DONT_EXECUTE VALID_COSTS ts tsk H_tsk_in_ts
    VALID_PREEMPTION_MODEL VALID_RTCT H0 VALID_ARRIVAL_CURVE
    RESPECTS_MAX_ARRIVALS
      (rs_jlfp_interference arr_seq sched)
      (rs_jlfp_interfering_workload arr_seq sched)
    ABSTRACT_WORK_CONSERVING H_sequential_tasks
    I_AND_W_SEQUENTIAL
    L BUSY_INTERVALS_BOUNDED (arm_sbf (*@$\Pi$@*) (*@$\Theta$@*)(*@$\nu$@*))
    INTRA_BOUNDED R SOL_SEQ_RS
    j ARR TSK).
\end{lstlisting}

\begin{lstlisting}[
 basicstyle=\fontsize{6}{7}\selectfont\ttfamily,
breaklines=false,
  columns=fullflexible,
  escapeinside={(*@}{@*)}
]
apply : uniprocessor_response_time_bound_restricted_supply_seq; 
  try done. 
\end{lstlisting}
\caption{Example illustrating the goal-driven lemma instantiation proving skill. The upper part shows the proof script using 
\texttt{exact}, i.e., explicitly providing all the needed preconditions and parameters. The lower part shows the proof script using \texttt{apply} combined with \texttt{try done}.}
\label{fig:goal-driven}
\end{figure}

Due to space limitations, we cannot elaborate on every proving skill in Table \ref{tab:proving_skill_summary}. In the following, we only give a detailed explanation of the \emph{goal-driven lemma instantiation} proving skill (the last row in Table \ref{tab:proving_skill_summary}), which addresses a problem specific to schedulability analysis using Prosa.
As mentioned in Section \ref{ss:overview}, Prosa has a highly structured hierarchy that facilitates lemma reuse across different task models. Applying such a lemma requires satisfying its preconditions, which themselves depend on further preconditions, eventually reaching the specific assumptions of the chosen task model. 
In these cases, the agent tends to use the \texttt{exact} tactic when applying such lemmas, which requires all preconditions and parameters to be explicitly supplied. The upper part of Figure \ref{fig:goal-driven} shows a real example we take from the traces for generating the proving skills. In this example, the lemma requires a long list of preconditions and parameters (lines~2 - 16) if using \texttt{exact} to apply this lemma.
It is very difficult for the LLM to precisely provide all the required preconditions and parameters. Even if the agent repeatedly guesses parameters, receives error messages from Rocq tools, and tries again, in most cases, it still can never find the correct combination. 
However, a simple solution exists for this problem: instead of using \texttt{exact} with all parameters, one can use the \texttt{apply} tactic with the lemma. This turns the preconditions and parameters into explicit goals, which are typically much easier to prove (many can be resolved automatically by built-in tactics). For instance, the lower part of Figure \ref{fig:goal-driven} shows the proof script that applies the lemma using \texttt{apply}, without providing the preconditions, and then using \texttt{try done} can close all the generated goals. This insight becomes a reusable proving skill.

Why does analyzing proof traces enable the LLM to generate useful skills? The traces contain rich diagnostic information: they show iterative attempts to solve a local problem, with clear success/failure outcomes, error messages, and the current proof state at each step. The LLM can observe how the agent fails repeatedly and how it eventually succeeds. The resulting proving skill thus encodes a practical solution that the LLM alone would not produce.

\begin{table*}[t]
\centering
\scriptsize
\setlength{\tabcolsep}{4pt}
\renewcommand{\arraystretch}{1.18}
\caption{
Rocq command classifications \cite{rocq-vernacextend-v91}.
}
\label{tab:vernac-classification-policy}

\begin{tabularx}{0.99\textwidth}{
    @{}
    >{\raggedright\arraybackslash}p{0.1\textwidth}
    >{\raggedright\arraybackslash}p{0.4\textwidth}
    >{\raggedright\arraybackslash}p{0.38\textwidth}
    >{\raggedright\arraybackslash}X
    @{}
}
\toprule
\textbf{Classification}
&
\textbf{Operational role}
&
\textbf{Representative commands}
&
\textbf{Decision}
\\
\midrule

\texttt{VtProofStep}
&
Advances or organizes the current proof state.
&
Tactic sentences, bullets, focusing commands and delimiters.
&
Allowed
\\

\texttt{VtQuery}
&
Queries without modifying the
interpretation state.
&
\texttt{Check}, \texttt{Print}, \texttt{Search}, \texttt{Locate}, and
\texttt{Show}.
&
Allowed
\\

\texttt{VtStartProof}
&
Initiates an interactive proof.
&
\texttt{Lemma}, \texttt{Theorem}, and proof-style \texttt{Definition}.
&
Forbidden
\\

\texttt{VtQed}
&
Terminates or discards the current interactive proof.
&
\texttt{Qed}, \texttt{Defined}, \texttt{Admitted}, and \texttt{Abort}.
&
Forbidden
\\

\texttt{VtProofMode}
&
Changes the proof mode used for subsequent proof commands.
&
Proof-mode switching commands, e.g.,
\texttt{Proof Mode "Ltac2"}.
&
Forbidden
\\

\texttt{VtMeta}
&
Represents STM, history-management, or proof-control commands.
&
\texttt{Back}, \texttt{Undo}, \texttt{Restart}, and reset commands.
&
Forbidden
\\

\texttt{VtSideff}
&
May modify the global environment, parser state, or interpretation state.
&
\texttt{Axiom}, \texttt{Parameter}, \texttt{Definition},
\texttt{Require}, \texttt{Notation}, \texttt{Set}/\texttt{Unset},
\texttt{Hint}, \texttt{Instance}, and \texttt{Declare ML Module}.
&
Forbidden
\\

\bottomrule
\end{tabularx}\label{t:classification}
\end{table*}

It is worth comparing proving skills with the decomposition guidelines. Both serve the same purpose: adding extra information to the LLM's prompt to improve its problem-solving ability. The differences lie in how they are obtained and used. Decomposition guidelines are crafted manually by human experts, while proving skills are generated by having the LLM analyze proof traces, followed by human review. We also attempted to use a similar trace-driven process to extract decomposition guidelines, but the results were poor. This is probably because the impact of a decomposition is non-local and diffuse; the LLM struggles to connect later proof failures to  decomposition choices and infer how the decomposition should be refined. 
Moreover, while proving skills are stored in skill files and loaded on demand, decomposition guidelines are placed directly in the agent's prompt. The current set of decomposition guidelines is small, so the benefit of making them load-on-demand is limited, but nothing prevents users from packaging them as on-demand skill files.

\subsection{Tools}
\label{subsec:external_tools}
In addition to the standard built-in tools in OpenCode, such as file reading and code search, we develop several extra tools to 
 assist the proof construction.

We integrate a suite of Rocq tools that supply fine-grained, context-aware feedback based on the current proof state.
Rocq provides a batch compiler and an interactive toplevel, i.e., a REPL (read--eval--print loop). \texttt{rocq-lsp} is a separate language server/plugin for Rocq. Our tools wrap these underlying interfaces.
We develop a compiler-backed tool that checks whole \texttt{.v} files and reports compilation status and errors, a toplevel-backed tool that answers queries about library objects, lemma types, assumptions, and definitions, and a \texttt{rocq-lsp}-backed tool that provides incremental diagnostics, current goals and local hypotheses.
Our integration follows OpenCode's standard way of providing tools. Each tool is a module with a call name, a brief description, and an input schema. The agent does not run Rocq directly. It selects a tool and supplies arguments, and the OpenCode runtime dispatches the call, using session configuration such as file paths and project roots. The output is written to files, which the OpenCode harness parses into structured results such as error locations, goals, and hypotheses. This gives the agent fine-grained proof-state feedback while avoiding fragile raw command-line handling.

We developed a structured proof-planning tool that enables the primary agent to externalize its decomposition as an explicit DAG before constructing a proof sketch that can  be compiled by Rocq.
Each node in the DAG records a Rocq subgoal, its dependencies, and optionally proposed library lemmas; terminal delegation candidates are materialized as annotated proof regions. The tool mechanically checks target-lemma consistency, acyclicity, and lemma applicability, and compiles the resulting sketch to ensure the subgoals can be combined. It also assigns each delegated subgoal a unique proof region, so a subagent can edit only its assigned region while preserving the surrounding proof and local context.
This tool is useful since, 
if a decomposition remains only in the primary agent's free-form reasoning, invalid decompositions may go unnoticed until subagents fail.

We also developed a tool to detect stagnation in proof construction. Since the Rocq compiler reports only the current proof state, it cannot determine whether repeated agent actions are advancing the proof or reproducing the same failure; without an independent progress assessment, the agent may keep calling the same tools, launching subagents on unchanged proof regions, or making superficial edits while wasting computation and tokens. At proof milestones, such as after the primary agent writes or substantively revises a proof sketch, after a proof region is updated, or after a subagent returns, the tool invokes Rocq, compares the new state with the previously recorded one, and records verified, structural, or diagnostic progress. It then uses these records to detect stagnation: if repeated tool calls or subagent attempts on an unchanged region produce no new evidence, it warns the agent and may temporarily block further attempts until new compiler evidence appears or the region’s subgoal, dependencies, or proof commands are substantively revised.

\subsection{Proof Auditing}\label{subsec:auditing}
When an agent declares a proof complete, we cannot treat that declaration, or even a successful compilation, as evidence that the proof has actually been written, since the agent may ``cheat'' during proof construction. For example, an agent might silently change the theorem statement, weaken a definition, or introduce an unproven axiom. This could still result in a successful compilation, yet not give a valid proof of the intended lemma. We therefore need an independent auditing step to decide whether the proof is valid.

The goal of proof auditing is to ensure two things: (1) the lemma to be proved is unchanged, and (2) the generated proof indeed proves the lemma rather than bypassing its proof. ProsaBuddy runs in a sandbox, an isolated workspace that limits which files the agent can read and write. We place the libraries the agent is allowed to use in its read-only scope, so that the agent can import from them but cannot modify or recompile the libraries themselves. Based on this, the auditing consists of the following steps.

First, we check whether the submitted \texttt{.v} file, which contains the lemma to be proved and its proof script, is accepted by Rocq. Concretely, this means that when the Rocq compiler processes the file, it is parsed and elaborated without errors, and the proof terms produced by the tactics are accepted by the Rocq kernel. In particular, if the proof script contains tactics that leave unfinished goals, such as \texttt{admit} or \texttt{give\_up}, the check fails at this step.

Second, we check that the proof body, i.e., the part after \texttt{Proof.} and before \texttt{Qed.}, contains only commands for advancing or querying the proof state. Table \ref{t:classification} shows the classification of Rocq commands. Among these classes, only the \texttt{VtProofStep} and \texttt{VtQuery} classes are allowed in the proof body; all other classes are forbidden. We use the \texttt{astdump} plugin (part of \texttt{rocq-lsp}) to extract the AST of the \texttt{.v} file and check whether there is an AST node corresponding to a command in a forbidden class within the proof body. If so, the check fails at this step.

Third, we verify that nothing outside the proof body has been changed, except for imports from trusted libraries at the very beginning. Specifically, before the agent starts working, we store a copy of the original \texttt{.v} file containing the target lemma. After the agent submits its \texttt{.v} file, we compare the ASTs of the two files to ensure that any changes outside the proof body are limited to imports from trusted libraries at the very beginning; otherwise, the check fails. Currently, we treat only MathComp and Prosa as trusted libraries and assume that importing them at the beginning of the \texttt{.v} file does not introduce risks of altering the lemma to be proved or enabling the agent to bypass the proof.

A proof submitted by the agent is deemed valid only if it passes all the steps above. Otherwise, the submitted proof is rejected and returned to the agent for reconstruction.
Note that the audit is performed in a different OS process, so any changes to the Rocq environment state made by the proving agents cannot affect the Rocq environment state used in the audit. The above auditing procedure is conservative in the sense that a proof file submitted by the agent is not necessarily invalid just because it failed the audit. Nevertheless, the soundness of our proof auditing, both its design and implementation, has not yet been formally verified. Establishing a formally verified proof auditing procedure will be our future work.

Besides our proof-auditing script itself, we rely on several trusted bases to declare a proof valid. The scope of the trusted base depends on the property being checked. For kernel-level proof validity, the trusted base is the Rocq kernel: it is the component that actually checks proof terms, and Rocq is designed so that nothing else needs to be trusted for soundness. ProsaBuddy's proof-auditing procedure, however, relies on additional components. 
Our assumed trusted base for auditing consists of the Rocq kernel, the Rocq compiler, the tactics allowed to be used, the trusted libraries, the \texttt{astdump} plugin, and our proof-auditing script. Trust to these components is needed because our audit does not inspect the proof term directly. It inspects the AST and checks whether the AST contains forbidden commands. The absence of such commands in the AST, however, does not by itself guarantee that the proof term submitted to the kernel is free of them. The AST is extracted from the source file by the \texttt{astdump} plugin, and the proof term is obtained by compiling the source file, during which the tactics are executed. To infer from the absence of forbidden commands in the AST that the proof term is also free of them, we have to trust the \texttt{astdump} plugin, the compiler, the tactics and the imported libraries, as well as our proof-auditing script. A bug in any of these components might cause our auditing procedure to overlook an invalid proof, even though the Rocq kernel itself remains sound. However, the tools that agents use while generating proofs need not be trusted: they are used only to provide information guiding agents to construct proofs, and the correctness of this information does not affect the auditing procedure.



\section{Evaluation}\label{experiments}

\subsection{Comparison with State-of-the-Art}
\label{subsec:comparison_with_stateoftheart}
We compare \ours\ with four recent LLM-based automated theorem proving systems for Coq/Rocq: PALM \cite{palm}, AutoRocq\cite{autorocq}, Cobblestone\cite{cobblestone} and RocqStar~\cite{rocqstar}. 
The interaction layer with Coq of PALM and Cobblestone relies on coq-serapi, which has only been maintained up to Coq 8.20. This dependency prevents PALM from being directly applied to the Rocq environment. Since our experiments use Rocq 9.0 for RocqStar and \ours, we ported the subset of coq-serapi interaction functionality used by PALM to Rocq 9.0 for a fair comparison.
We also attempted to evaluate COPRA~\cite{copra}, another recent LLM-based Coq/Rocq proving agent, but COPRA failed on all benchmark tasks in our environment; therefore, we do not include its results in Table~\ref{tab:results_notprosa}. We investigated the failures and found that, although COPRA claims support for both Lean and Coq, the released GitHub implementation leaves several Coq-environment connection functions unimplemented. 
We did not compare with Rango~\cite{rango} because RocqStar has already included a direct comparison with Rango and reported better performance \cite{rocqstar}. Therefore, we only choose the better one as the representative baseline in our experiments. {We did not compare with CoqPilot because RocqStar is a follow-up system that extends the CoqPilot pipeline with a more advanced agentic architecture \cite{coqpilot, rocqstar}.
We did not compare with ADAPT\cite{adaptive}, QUARRY\cite{quarry}, ReCent-Prover~\cite{recentprover} and the work in \cite{liu2024llm} and \cite{lu2025clarifying} because they are not yet publicly available.}
Moreover, we also include the original OpenCode in the comparison. For all systems, we use GPT-5.4 as the backbone LLM. All compared systems have access to the full Prosa library.

\begin{table*}[htbp]
\label{tab:bench}
  \centering
  \caption{Information of the Lemma in our Benchmark} 
  \label{tab:bench}
  \tiny 
  \setlength{\tabcolsep}{2.2pt} 
  \begin{tabular}{@{}ccccccccccccccccccccccccccc@{}} 
    \toprule 
Index  & \text{\#}1 & \text{\#}2 & \text{\#}3 & \text{\#}4 & \text{\#}5 & \text{\#}6 & \text{\#}7 & \text{\#}8 & \text{\#}9 & \text{\#}10 &
    \text{\#}11 & \text{\#}12 & \text{\#}13 & \text{\#}14 & \text{\#}15 & \text{\#}16 & \text{\#}17 & \text{\#}18 & \text{\#}19 & \text{\#}20 & \text{\#}21 & \text{\#}22 & \text{\#}23 & \text{\#}24   \\
    \midrule 

    Source & \makecell{ \cite{bertogna2005}  \\ Lem 3} & \makecell{ \cite{bertogna2005}  \\ Lem 4} & \makecell{ \cite{bertogna2005}  \\ Thm 6} & \makecell{\cite{bertogna2007} \\ Thm 1} & \makecell{\cite{bertogna2007} \\ Thm 2} & \makecell{\cite{bertogna2007} \\ Thm 3} & \makecell{\cite{bertogna2007} \\ Thm 4}  & \makecell{\cite{guan2009} \\ Lem 1} &  \makecell{\cite{guan2009} \\ Thm 1 } &  \makecell{\cite{guan2009} \\ Lem 1-1} & 
    \makecell{\cite{guan2009} \\ Lem 1-2} & 
    \makecell{\cite{guan2009} \\ Lem 2-1} & \makecell{\cite{guan2009} \\ Lem 2-2} &\makecell{\cite{guan2009} \\ Lem 3} & \makecell{\cite{guan2009} \\ Lem 4} & \makecell{\cite{guan2009} \\ Lem 5} & 
    \makecell{\cite{guan2009} \\ Thm 1} &
    \makecell{\cite{guan2009} \\ Thm 2} & 
    \makecell{\cite{sun2014} \\ Thm 3} & \makecell{\cite{sun2014} \\ Lem 4} & \makecell{\cite{sun2014} \\ Lem 5} & 
    \makecell{\cite{baruah2015} \\ Lem 18.1} & \makecell{\cite{baruah2015} \\ Lem 18.6} & \makecell{\cite{2013ecrts} \\ Lem 8 }    \\    
    \midrule 
    Dependency & -  & \text{\#}1  & \text{\#}1, \text{\#}2 & - & - & \text{\#}3 & - & - & \text{\#}8 & - &
   - & \text{\#}10, \text{\#}11 & \text{\#}10, \text{\#}11 & \text{\#}12, \text{\#}13 & \text{\#}14 & - & \text{\#}15 & \text{\#}16 & \text{\#}21 & - & \text{\#}20 & - & \text{\#}22 & -    \\   

    \midrule 
    
   num. of Tactics   & 70-110 &160-200 & - & 180-300 & 10-40 & 70-90 &10-40 &60-80 & 30-70 &200-220 &
    150-200 & 40-60 & -& - &15-30 & - & 10-20 & 100-150& 20-80& - & - & 15-90 & - & 60-80&   \\
    
        \bottomrule 
     \end{tabular}

~\\
\scriptsize

{\raggedright 
Note: 
\enquote{Dependency} represents which other lemma in this table are prerequisites for this
lemma. 
\enquote{\# of Tactics} represents the \emph{range} of the number of tactics in the proof generated by different systems in the following evaluation experiments (different systems often generate proofs with different numbers of tactics). For those lemmas/theorems which are not proved by any system, the number is denoted by \enquote{-}.
\par}
\end{table*}   

The evaluation is conducted on a benchmark that we created.
The benchmark contains $24$ lemma formalizations in Rocq, which are not included in the existing Prosa library. These lemmas are drawn from five research papers \cite{bertogna2005,bertogna2007,guan2009,2013ecrts,sun2014} and one book \cite{baruah2015} in real-time scheduling theory, and we manually formalized them in Rocq. 
This ensures that the benchmark has no overlap with the \enquote{training} lemmas used to automatically generate the proving skills, as introduced in Section~\ref{subsec:proving_skills}.
The sources of these lemmas are listed in Table~\ref{tab:bench}.
All lemmas concern global multiprocessor scheduling. 
Note that, in Table~\ref{tab:bench},  case \text{\#}10 and \text{\#}11 (and also \text{\#}12 and \text{\#}13) come from manually splitting a single lemma into separate smaller lemmas. We do this when the original lemma states multiple independent results at once. This does not change the mathematical content of the original result. 
There is a compatibility issue: Prosa 0.4 includes foundations for global multiprocessor scheduling, but the latest version, Prosa 0.6, does not have this part. Prosa 0.6 is based on Rocq, whereas Prosa 0.4 is based on Coq, and Rocq and Coq have subtle syntactic differences. Therefore, we manually ported the global multiprocessor scheduling part in Prosa 0.4 from Coq to Rocq. 
 

{We use the relevant portion
of the corresponding research paper as the human-written proof hint input.
Specifically, this input includes the target theorem or lemma
itself, its full human-written proof, as well as any supporting
definitions, notation, and variables that are referenced within
the proof. AutoRocq, OpenCode and ProsaBuddy accept this input through
their existing interfaces. However, PALM, Cobblestone, and RocqStar do not originally provide such an interface. PALM and
Cobblestone operate through their generate-and-repair
workflows, which take the formal lemma and the current proof
state as inputs and iteratively attempt to construct a valid
proof. Neither PALM nor Cobblestone accepts additional
text inputs. RocqStar retrieves
formal definitions and theorems from a structured Rocq library,
but it does not accept free-text descriptions as input.
To provide the same human-written proof hint to all systems, we augmented PALM, Cobblestone, and RocqStar to prepend the corresponding hint to their system prompts. All
comparison results reported for PALM, Cobblestone, and RocqStar
are obtained from their augmented version with human-written hints.}

Table \ref{tab:results_notprosa} shows the comparison results.
Each column in the table is the lemma index in the benchmark. In each cell, \ding{52} represents that the lemma is successfully proved, while \ding{56} represents the agent declares a failure to prove this lemma and stops.
The number under \ding{52} represents the total amount of token consumption proving this lemma in the unit of $1$K tokens (counting both input tokens and output tokens). $\bm{\infty}$ represents that the agent has not yet proved the lemma before reaching the timeout threshold. In our experiments, we set the timeout threshold to be $50,000$K tokens. 
The timeout is measured by the number of tokens consumed rather than by actual time consumption, since variations in time delay depend heavily on many factors unrelated to the model inference, such as network latency, server load, or local hardware restrictions. In contrast, the token count provides a more stable metric that better reflects the amount of LLM inference performed.
To give readers a rough sense of scale, in most cases \ours\ takes between $3$ and $5$ hours to reach the $50,000$K‑token threshold. Besides the total token consumption, the KV-cache hit rate is also a key factor determines the proving efficiency and cost. The KV-cache hit rate of \ours\ is on average about $90\%$ in our experiments\footnote{The KV-cache hit rate also depends on the LLM serving provider's KV-cache resource allocation policy. For example, the KV-cache hit rate will drop if less KV-cache space is allocated to each request session. Therefore, this KV-cache hit rate result may not be reproducible.}.

To reduce time and cost, we omit some experiments, which are denoted by \enquote{--} in 
Table \ref{tab:results_notprosa}.
This is because \ours\ shows much better performance
than all the compared systems, so
we infer that for the cases where \ours\ failed, all the other compared system will also very likely fail, so we omit these experiments.

The experiment results show that \ours\ significantly outperforms the compared systems. Out of $24$ lemmas, {PALM and Cobblestone only succeed $1$ lemma, AutoRocq succeeds $2$, RocqStar and OpenCode succeed $3$, while \ours\ succeed $17$}. For the one lemma PALM and Cobblestone succeed, they use many fewer tokens. This is because of the way they work:
they invoke the LLM for a specific number of times, and detailed error corrections rely on automated proof generation tools such as CoqHammer\cite{coqhammer}. This differs from other agent-based systems, which conduct iterative dialogues with the LLM for both proof generation and error correction.

Our benchmark targets lemmas that are not yet included in Prosa but are closely connected to the foundations provided by Prosa. Their proofs rely on existing Prosa definitions and basic lemmas. This selection is intentional: lemmas that are largely unrelated to the current Prosa foundation would require \ours\ to invent new definitions and many intermediate lemmas from scratch, which is beyond its current capability.
Accordingly, the experimental results should not be interpreted as showing that \ours\ can prove real-time scheduling theorems in full generality. They instead evaluate whether \ours\ can extend an existing mechanized library by proving new lemmas when the relevant definitions and many auxiliary results are already available. This reflects the positive feedback loop we aim to support: as Prosa grows to cover more real-time scheduling topics, \ours\ can benefit from the richer foundation and help prove further lemmas.

\begin{table*}[htbp]
  \centering
  \caption{Results of the Comparison Experiments and Ablation Study with our Built Benchmark}
  \label{tab:results_notprosa}
  \tiny
  \setlength{\tabcolsep}{3pt}
  \begin{tabular}{@{}cccccccccccccccccccccccccccc@{}}
    \toprule
    Index & \text{\#}1 & \text{\#}2 & \text{\#}3 & \text{\#}4 & \text{\#}5 & \text{\#}6 & \text{\#}7 & \text{\#}8 & \text{\#}9 & \text{\#}10 &
    \text{\#}11 & \text{\#}12 & \text{\#}13 & \text{\#}14 & \text{\#}15 & \text{\#}16 & \text{\#}17 & \text{\#}18 & \text{\#}19 & \text{\#}20 & \text{\#}21 & \text{\#}22 & \text{\#}23 & \text{\#}24 &
    \makecell{num. of \\ success} \\
    \midrule

    PALM & \ding{56} & \ding{56} & - & \ding{56} & \makecell{\ding{52} \\ 3} & \ding{56} & \ding{56} & \ding{56} & \ding{56} & \ding{56} &
    \ding{56} & \ding{56} & - & - & \ding{56} & - & \ding{56} & \ding{56} & \ding{56} & - & - & \ding{56} &
    - & \ding{56} & \textbf{1}\\

    RocqStar & \ding{56} & \ding{56} & - & \ding{56} & \makecell{\ding{52} \\ 23} & \ding{56} & \ding{56} & \ding{56} & \ding{56} & \ding{56} &
    \ding{56} & \ding{56} & - & - & \ding{56} & - & \makecell{\ding{52} \\ 587} & \ding{56} & \makecell{\ding{52} \\ 1,486} &
    - & - & \ding{56} & - & \ding{56} & 3\\

     AutoRocq & \ding{56} & \ding{56} & - & \ding{56} & \makecell{\ding{52} \\ 6} & \ding{56} & \ding{56} & \ding{56} & \ding{56} & \ding{56} &
    \ding{56} & \ding{56} & - & - & \makecell{\ding{52} \\ 1176} & - & \ding{56}  & \ding{56} & \ding{56} &
    - & - & \ding{56} & - & \ding{56} & 2\\

    Cobblestone & \ding{56} & \ding{56} & - & \ding{56} & \makecell{\ding{52} \\ 2} & \ding{56} & \ding{56} & \ding{56} & \ding{56} & \ding{56} &
    \ding{56} & \ding{56} & - & - & \ding{56} & - & \ding{56}  & \ding{56} & \ding{56} &
    - & - & \ding{56} & - & \ding{56} & 1\\

    \makecell{OpenCode} & \ding{56} & \ding{56} & - & \ding{56} & \makecell{\ding{52} \\ 247} & \ding{56} & \ding{56} & \ding{56} &
    \makecell{\ding{52} \\ 1,083} & \ding{56} & \ding{56} & \ding{56} & - & - & \ding{56} & - &
    \makecell{\ding{52} \\ 500} & \ding{56} & \ding{56} & - & - & \ding{56} & - & \ding{56} & 3\\

    \midrule

    PB-w/o-S & \makecell{\ding{52} \\ 25,410} & $\bm{\infty}$ & - & \makecell{\ding{52} \\ 23,440} &
    \makecell{\ding{52} \\ 206} & \makecell{\ding{52} \\ 3,760} & $\bm{\infty}$ & \makecell{\ding{52} \\ 21,540} &
    \makecell{\ding{52} \\ 535} & \ding{56} &
    \makecell{\ding{52} \\ 45,185} & $\bm{\infty}$ & - & - & \makecell{\ding{52} \\ 4,084} & - &
    \makecell{\ding{52} \\ 601} & \makecell{\ding{52} \\ 2,398} &
    \makecell{\ding{52} \\ 896} & - &
    - & \makecell{\ding{52} \\ 29,129} & - & \ding{56} & 12 \\

    PB-w/o-D & \makecell{\ding{52} \\ 23,056} & \makecell{\ding{52} \\ 20,858} & - & \makecell{\ding{52} \\ 11,296} &
    \makecell{\ding{52} \\ 186} & \makecell{\ding{52} \\ 11,481} & \makecell{\ding{52} \\ 2,617} & \ding{56} &
    \makecell{\ding{52} \\ 456} & $\bm{\infty}$ &
    \makecell{\ding{52} \\ 30,334} & $\bm{\infty}$ & - & - & \makecell{\ding{52} \\ 11,095} & - &
    \makecell{\ding{52} \\ 616} & \makecell{\ding{52} \\ 8,798} &
    \makecell{\ding{52} \\ 479} & - & - & \makecell{\ding{52} \\ 8,680} & - & \makecell{\ding{52} \\ 29,137} & 14 \\

    PB-w/o-P & \makecell{\ding{52} \\ 37,125} & \ding{56} & -& \ding{56} &
    \makecell{\ding{52} \\ 205} & \makecell{\ding{52} \\ 30,139} & \makecell{\ding{52} \\ 47,333} & \ding{56} &
    \makecell{\ding{52} \\ 6,511} & \makecell{\ding{52} \\ 35,156} &
    \makecell{\ding{52} \\ 24,619} & \ding{56} & - & - & \makecell{\ding{52} \\ 3,317} & - &
    \makecell{\ding{52} \\ 2,177} & \makecell{\ding{52} \\ 9,354} &
    \makecell{\ding{52} \\ 1,625} & - & - & \makecell{\ding{52} \\ 27,604} & - & \ding{56} & 12 \\

    PB-w/o-H & \makecell{\ding{52} \\ 36,340} & $\bm{\infty}$ & - & \makecell{\ding{52} \\ 11,238} &
    \makecell{\ding{52} \\ 358} & \makecell{\ding{52} \\ 3,501} & \makecell{\ding{52} \\ 1,257} & \makecell{\ding{52} \\ 16,550} &
    \makecell{\ding{52} \\ 1,398} & \makecell{\ding{52} \\ 41,562} &
    \makecell{\ding{52} \\ 31,716} & \ding{56} & - & - & \makecell{\ding{52} \\ 23,906} & - &
    \makecell{\ding{52} \\ 1,321} & \makecell{\ding{52} \\ 5,096} &
    \makecell{\ding{52} \\ 1,833} & - & - & \makecell{\ding{52} \\ 21,157} & - & \makecell{\ding{52} \\ 9,278} & 15 \\

    \midrule

    \textbf{\ours} & \makecell{\ding{52} \\ 48,819} & \makecell{\ding{52} \\ 42,027} & $\bm{\infty}$ & \makecell{\ding{52} \\ 13,332} &
    \makecell{\ding{52} \\ 281} & \makecell{\ding{52} \\ 1,295} & \makecell{\ding{52} \\ 2,077} & \makecell{\ding{52} \\ 13,077} &
    \makecell{\ding{52} \\ 1,888} & \makecell{\ding{52} \\ 24,330} &
    \makecell{\ding{52} \\ 36,004} & \makecell{\ding{52} \\ 38,749} & \ding{56} & \ding{56} & \makecell{\ding{52} \\ 6,736} & \ding{56} &
    \makecell{\ding{52} \\ 3,908} & \makecell{\ding{52} \\ 2,229} &
    \makecell{\ding{52} \\ 307} & \ding{56} & $\bm{\infty}$ & \makecell{\ding{52} \\ 31,614} & \ding{56} & \makecell{\ding{52} \\ 13,557} & 17 \\
    \bottomrule
  \end{tabular}
  ~\\
  \scriptsize
  {\raggedright 
~~~~~~~~~~ Note: The number under each \ding{52} represents the amount of token consumption to prove this lemma, which is in thousands of tokens.\par}
\end{table*}

{%
The results in Table~\ref{tab:results_notprosa} report only a single run for each case by each system. Since LLM generation is inherently stochastic, repeating the evaluation multiple times per case would yield more robust conclusions. However, conducting such repeated runs for all cases would incur prohibitive time and costs. As a compromise, we select $6$ cases to further conduct multi-run evaluations.
These cases are selected
based on the token consumption of ProsaBuddy in the above experiments, covering different difficult levels: \text{\#}15 and  \text{\#}17 as the low-difficulty group, \text{\#}8 and  \text{\#}22 as the medium-difficulty  group, and \text{\#}1 and  \text{\#}2 as the high-difficulty group. For each case, we run each system $5$ times under the same configuration. $x/5$ indicates that the system succeeds $x$ times out of the five runs. For $n=5$ runs with $x$ successes, we compute $\mathrm{pass@}k = 1 - \binom{n-x}{k} / \binom{n}{k}$. Token counts are also reported in thousands. The numbers in the `Range' and `Average' rows represents the token consumption range and the average token consumption of the $5$ runs.} 

\begin{table}[t]
  \centering
  \caption{{Results of Five Runs}}
  \label{tab:multiple}
  \tiny
  \setlength{\tabcolsep}{2.2pt}
  \renewcommand{\arraystretch}{1.15}
  {%
  \resizebox{\columnwidth}{!}{%
  \begin{tabular}{lcccccc}
    \toprule
   Difficulty & \multicolumn{2}{c}{Low} &
      \multicolumn{2}{c}{Medium} &
      \multicolumn{2}{c}{High} \\
    \cmidrule(lr){2-3}\cmidrule(lr){4-5}\cmidrule(lr){6-7}
    Index & \text{\#}15& \text{\#}17 & \text{\#}8 & \text{\#}22 & \text{\#}1 & \text{\#}2 \\
    \midrule
    PALM & 0/5 & 0/5 & 0/5 & 0/5 & 0/5 & 0/5 \\
    RocqStar & 0/5 & 1/5 & 0/5 & 0/5 & 0/5 & 0/5 \\
    AutoRocq & 2/5& 0/5 & 0/5 & 0/5 & 0/5 & 0/5\\
    Cobblestone & 0/5 & 0/5 & 0/5 & 0/5 & 0/5 & 0/5\\
    OpenCode & 0/5 & 2/5 & 0/5 & 0/5 & 0/5 & 0/5 \\
    \textbf{\ours} & 5/5 & 5/5 & 5/5 & 4/5 & 4/5 & 4/5 \\
    \midrule
    pass@1 & 1 & 1 & 1 & 0.8 & 0.8 & 0.8 \\
    pass@3 & 1 & 1 & 1 & 1 & 1 & 1 \\
     Range & 6,736--19,087 & 1,748-6,457 &
      13,077--38,724 & 22,890--31,614 & 29,168--54,525 &
      35,546--64,957 \\
    Average & 11,471& 3,669 & 25,220 &
      28,120 & 44,171 & 50,630 \\
    \bottomrule
  \end{tabular}%
  }
  }
  \vspace{1mm}
\end{table}

{%
The results in Table~\ref{tab:multiple} show that, across the $6$ selected cases, \ours\ achieves $27$ successful proofs out of $30$ total runs. It succeeds in all $5$ runs on Cases \text{\#}8, \text{\#}15, and \text{\#}17, and in $4$ of $5$ runs on Cases \text{\#}1, \text{\#}2, and \text{\#}22. These repeated runs confirm that both proof outcomes and token consumption vary to some extent across individual executions. However, this variation does not alter the main comparative picture: \ours\ attains at least four successes out of five runs on every selected case, whereas PALM yields no successful runs and OpenCode succeeds in only $2$ of its $30$ runs. Furthermore, the estimated pass@3 for \ours\ is 1.0 for all $6$ cases.}

\begin{table}[htbp]
\caption{Results with Lemmas Selected from Prosa}
\centering
\tiny
\renewcommand{\arraystretch}{1.6}
\begin{tabular}{cccccccc}
\toprule
 & L1 & L2 & L3 & L4 & L5 & L6 & L7 \\
num.  of tactics& 1-20 & 21-40 & 41-60 & 61-80 & 81-100 & 101-150 & $>$ 150 \\
\hline
PALM & 1/10 & 0/10 & - & - & - & - & - \\
RocqStar & 7/10 & 2/10 & - & - & - & - & - \\
AutoRocq & 5/10 & 1/10 & - & - & - & - & - \\
Cobblestone & 1/10 & 0/10 & - & - & - & - & - \\
OpenCode & 4/10 & 0/10 & - & - & - & - & - \\
    \midrule 
\textbf{\ours} & 10/10 & 
10/10 & 8/10 & 9/10 & 9/10 & 6/10 & 1/5 \\

\bottomrule
\end{tabular}
~\\
\footnotesize
Note: \enquote{\# of tactics} represent the number of tactics used in its proof.
\label{tab:prosaexp}
\end{table}

We further conduct evaluations using more lemmas selected from Prosa. These lemmas are categorized into seven levels (L1–L7) based on proof length, measured by the number of tactics used in the proof. Each level contains $10$ lemmas, except L7: Prosa has $12$ L7 lemmas in total, among which we selected $5$ as training lemmas and $5$ as evaluation lemmas for this evaluation, with the evaluation lemmas strictly distinct from the training lemmas used in Section \ref{subsec:proving_skills}.
Unlike the our build benchmark, this experiment does not provide the agent with human-written proof hints. This is because many lemmas in Prosa are library-internal results and do not have corresponding human-written proofs in the literature.

We observe that Prosa contains many families of similar proofs.
For example, the same lemma is established under several related scheduling models, such as fixed-priority preemptive, non-preemptive, and limited-preemptive scheduling.
Although these are technically different lemmas, their statements and proof structures can be very similar.
In such cases, an agent with access to the full library may solve the target lemma largely by copying or lightly adapting an existing proof, rather than by generating the proof from the target statement and local context.
This would make the evaluation less interesting.
To solve this problem, for each lemma, we provide only the minimal Prosa slice required to compile the target file, computed from the transitive dependency closure of its required modules. In this way, the agent has enough formal context to prove the benchmark lemma, but it cannot search other parts of Prosa for the original proof or a similar proof.

Similar to above experiments, we omit the experiments in L3-L7 for the compared systems, which are denoted by \enquote{--} in 
Table \ref{tab:prosaexp}.
This is because, if a baseline shows poor ability to solve lemmas in lower-level groups, there is no reason to expect better performance on higher levels.
As PALM, RocqStar, Autorocq, Cobblestone and OpenCode solved zero or only a very small number of lemmas at L2.
Therefore, we did not further evaluate them on L3 and above.
The experiment results with lemmas selected from Prosa confirm the superiority of \ours. The results also show that \ours\ still can successfully prove non-trivial lemmas even without human-written proof hints.
\begin{table*}[htbp]
  \centering
  \caption{Experiment Results with Different Backbone LLMs}
  \label{tab:diff_llm}
  \tiny
  \setlength{\tabcolsep}{3pt}
  \begin{tabular}{@{}cccccccccccccccccccccccccccc@{}}
    \toprule
Index    & \text{\#}1 & \text{\#}2 & \text{\#}3 & \text{\#}4 & \text{\#}5 & \text{\#}6 & \text{\#}7 & \text{\#}8 & \text{\#}9 & \text{\#}10 &
    \text{\#}11 & \text{\#}12 & \text{\#}13 & \text{\#}14 & \text{\#}15 & \text{\#}16 & \text{\#}17 & \text{\#}18 & \text{\#}19 & \text{\#}20 & \text{\#}21 & \text{\#}22 & \text{\#}23 & \text{\#}24 &
    \makecell{num. of \\ success} \\
    \midrule

    GPT-5.4 & \makecell{\ding{52} \\ 48,819} & \makecell{\ding{52} \\ 42,027} & $\bm{\infty}$ & \makecell{\ding{52} \\ 13,332} & \makecell{\ding{52} \\ 281} & \makecell{\ding{52} \\ 1,295} & \makecell{\ding{52} \\ 2,077} & \makecell{\ding{52} \\ 13,077} &
    \makecell{\ding{52} \\ 1,888} & \makecell{\ding{52} \\ 24,330} &
    \makecell{\ding{52} \\ 36,004} & \makecell{\ding{52} \\ 38,749} & \ding{56} & \ding{56} & \makecell{\ding{52} \\ 6,736} & \ding{56} &
    \makecell{\ding{52} \\ 3,908} & \makecell{\ding{52} \\ 2,229} &
    \makecell{\ding{52} \\ 307} & \ding{56} & $\bm{\infty}$ & \makecell{\ding{52} \\ 31,614} & \ding{56} & \makecell{\ding{52} \\ 13,557} & 17 \\

    DeepSeek-v4-flash & $\bm{\infty}$ & $\bm{\infty}$ & - & $\bm{\infty}$ & \makecell{\ding{52} \\ 959} & \makecell{\ding{52} \\ 32,528} & \makecell{\ding{52} \\ 7,406} & \ding{56} &
    \makecell{\ding{52} \\ 4,013} & $\bm{\infty}$ &
    $\bm{\infty}$ & $\bm{\infty}$ & - & - & \makecell{\ding{52} \\ 6,205} & - &
    \makecell{\ding{52} \\ 2,262} & \makecell{\ding{52} \\ 18,405} &
    \makecell{\ding{52} \\ 5,624} & - & - & \makecell{\ding{52} \\ 20,245} & - & $\bm{\infty}$ & 9 \\

    DeepSeek-v4-pro & $\bm{\infty}$ & $\bm{\infty}$ & - & $\bm{\infty}$ & \makecell{\ding{52} \\ 2,201} & $\bm{\infty}$ & \makecell{\ding{52} \\ 8,749} & \ding{56} &
    \makecell{\ding{52} \\ 4,247} & \ding{56} &
    \ding{56} & \ding{56} & {-} & {-} & \makecell{\ding{52} \\ 20,898} & {-} &
    \makecell{\ding{52} \\ 768} & \ding{56} &
    \makecell{\ding{52} \\ 4,567} & {-} & {-} & \ding{56} & - & \ding{56} & 6 \\

    Gemini-3.1-flash-lite & \ding{56} & $\bm{\infty}$ & {-} & $\bm{\infty}$ & \makecell{\ding{52} \\ 1,889} & $\bm{\infty}$ & $\bm{\infty}$ & $\bm{\infty}$ &
    \ding{56} & $\bm{\infty}$ &
    \ding{56} & \ding{56} & {-} & {-} & $\bm{\infty}$ & {-} &
    $\bm{\infty}$ & \ding{56} &
    \makecell{\ding{52} \\ 10,355} & {-} & {-} & \ding{56} & {-} & $\bm{\infty}$ & 2 \\

    GPT-oss-120B & \ding{56} & \ding{56} & - & \ding{56} & \makecell{\ding{52} \\ 1,192} & \ding{56} & \ding{56} & \ding{56} &
    \ding{56} & \ding{56} &
    \ding{56} & \ding{56} & - & - & \ding{56} & {-} &
    \ding{56} & \ding{56} &
    \ding{56} & - & - & \ding{56} & - & \ding{56} & 1 \\
    \bottomrule
  \end{tabular}
\end{table*}

\subsection{Ablation Study}

We conduct ablation experiments to evaluate the contribution of different components in \ours, the result of which is also reported in Table~\ref{tab:results_notprosa}.

\textbf{PB-w/o-S}. This variant removes the proving skills from \ours. It proves $12$ lemmas, fewer than the full \ours\ does, demonstrating the effectiveness of the proving skills developed in Section \ref{subsec:proving_skills}.

\textbf{PB-w/o-D}. This variant removes the lemma decomposition mechanism and instead uses only the primary agent to prove a lemma. PB-w/o-D proves $14$ lemmas. The gap between PB-w/o-D and \ours\ mainly appears on harder cases, i.e., lemmas on which \ours\ spends more tokens. Token counts reveal a trade-off: \ours\ sometimes spends more tokens because planning and delegation introduce extra interaction cost, but this overhead enables proofs that the single-agent variant cannot complete.

\textbf{PB-w/o-H}. This variant removes the human-written proof hints from \ours, i.e., it does not read the corresponding paper content. PB-w/o-H proves $15$ lemmas. The two cases where \ours\ succeeds but PB-w/o-H does not are both relatively difficult lemmas. This suggests that for simple cases, the formal context in Prosa, including definitions, existing lemmas, and proof states, already provides substantial guidance for proof construction. However, for difficult lemmas, the human-written proof hints play a significant role.

\textbf{PB-w/o-P}. This variant removes access to the full Prosa library and only provides the minimal dependent files needed to state the target lemmas. As a result, it can use basic definitions and low-level lemmas (related to the basic properties of the definitions) included in the dependent files, but cannot directly observe or reuse higher-level auxiliary lemmas in Prosa. PB-w/o-P proves $12$ lemmas. This shows that a full library context is important for extending an existing mechanized development. Without it, the agent needs to rediscover intermediate facts that are already available in Prosa, which signiciantly increase the difficulty to complete the proof. For most cases where \textbf{PB-w/o-P} succeeds, it consumes more token than \ours, which witnesses such extra overhead even if the lemma is still proved. 

\subsection{Different Backbone LLMs}

We further evaluate how the choice of different backbone LLM affects the performance of \ours. The results are reported in Table~\ref{tab:diff_llm}. In these experiments, all components of \ours\ are held constant except the backbone LLM. 
From the results we can see that the backbone LLM has a substantial impact on proof success: with the most powerful model, GPT-5.4, \ours\ proves 
17 lemmas, compared to 9 with DeepSeek-v4-flash, 
6 with DeepSeek-v4-pro, 2 with Gemini-3.1-flash-lite, and 1 with GPT-oss-120B.
Unexpectedly, DeepSeek-v4-flash outperforms DeepSeek-v4-pro. We currently do not have a clear explanation for this anomaly and will investigate its cause in future work.
Overall, these results show that the agent framework, augmented external knowledge, and feedback from Rocq tools are not sufficient on their own; the reasoning capability of the backbone LLM is crucial for the performance of \ours.
We also observe that, on the subset of lemmas the weaker models can prove, they tend to consume more tokens on average than GPT-5.4, with a few exceptions. Similar to above, the mark \enquote{--} in Table \ref{tab:diff_llm} denotes the omitted experiments to save time and cost, as for the difficult cases GPT-5.4 failed, the other weaker LLMs likely will also fail.

\section{Discussions and Conclusions}\label{conclusion}
In this work, we present
ProsaBuddy, an LLM-based agent system designed to lower
the effort needed to develop mechanized real-time schedulability proofs. 
Evaluations show that \ours\ can successfully prove many non-trivial lemmas. However, this is only a preliminary attempt in this direction; it has many limitations and ample room for further exploration.

Although in this paper we introduce and evaluate \ours\ as a standalone system that proves lemmas end-to-end, our ultimate goal is to build \ours\ as a highly interactive tool. When a human expert intervenes in the proving process and provides guidance at points where \ours\ gets stuck, \ours\ becomes more effective and useful. We did not systematically evaluate \ours\ with a human in the loop, because designing an objective and reproducible evaluation methodology for such interaction would be difficult. Nevertheless, we conducted some ad‑hoc interactive experiments. For example, for cases $\text{\#}13$ and $\text{\#}14$ in Table~\ref{tab:results_notprosa}, which \ours\ could not prove alone, a researcher with expertise in real‑time scheduling theory worked interactively with \ours, injecting additional human instructions when \ours\ was stuck or actively requested support. In this way, both lemmas were successfully proved in $2$ hours and $3$ hours, respectively. For comparison, a second researcher with a similar background independently proved them fully manually in Rocq, requiring $14$ hours and $17$ hours, respectively.

A critical issue in automated theorem proving is the formalization of the lemmas themselves. This is especially important for schedulability lemmas, whose descriptions are lengthy and involve many definitions and hypotheses. Correct formalization is the foundation for the proof work: if a lemma is incorrectly formalized, its proof is meaningless. Therefore, the lemma must be formalized very carefully, a task that itself demands considerable effort. Assisting researchers during this formalization phase is thus also valuable. However, while lemma proving benefits from an automatic verifier (Rocq) that can check correctness, formalization lacks such a verifier. It is impossible to fully automate the checking of whether a formalization precisely captures the human's intended meaning. Our future work will investigate how to utilize an LLM-based agent to assist the formalization of schedulability analysis lemmas.

The decomposition guidelines and proving skills introduced in Section \ref{s:prosabuddy} are only a small subset of what can be added to improve the agent’s capabilities. We expect that ProsaBuddy users will develop their own skills to enhance the agent, and we look forward to an ecosystem in which users share their skills to collectively improve the system. Such an ecosystem would also create a feedback loop in which successful skills are refined, reused, and benchmarked across projects, accelerating collective improvement. On the other hand, these guidelines and skills are tailored to the capabilities of today’s LLMs. Given how rapidly LLMs are advancing, future LLMs may have much stronger reasoning abilities, rendering many previously useful skills obsolete.

\ours\ has the potential to be used more broadly. Proofreader \cite{proofreader} is an agentic system that helps researchers and authors find flaws in their own proofs. It actively seeks counterexamples and errors, without aiming to establish formal correctness. Given these complementary strengths, future work could explore how \ours\ and other tools like Proofreader might jointly accelerate the discovery, refinement, and validation of new theorems.

\bibliographystyle{IEEEtran}
\bibliography{sections/reference}

\end{document}